\documentclass[12pt,preprint]{aastex}

\newcommand{\vv}[1]{\textbf{#1}}

\newcommand{\Msun}{M_{\odot}}
\newcommand{\Ma}{M_{\rm a}}
\newcommand{\Mc}{M_{\rm c}}

\begin{document}
\title
{
Frequency spectrum of gravitational radiation from
global hydromagnetic oscillations
of a magnetically confined mountain
on an accreting neutron star
}

\author
{D. J. B. Payne and A. Melatos}
\affil
{School of Physics, University of Melbourne,
Parkville, VIC 3010, Australia.
}
\email
{dpayne@physics.unimelb.edu.au}

\begin{abstract}
Recent time-dependent, ideal-magnetohydrodynamic (ideal-MHD)
simulations of polar magnetic burial
in accreting neutron stars have
demonstrated that stable, magnetically confined mountains 
form at the magnetic poles,
emitting gravitational waves
at $f_{*}$ (stellar spin frequency) and $2 f_{*}$.
Global MHD oscillations of the mountain,
whether natural or stochastically driven, act to
modulate the gravitational wave signal,
creating broad sidebands
(full-width half-maximum $\sim 0.2f_*$)
in the frequency spectrum
around $f_{*}$ and $2 f_{*}$.
The oscillations can enhance the signal-to-noise
ratio achieved by a
long-baseline interferometer
with coherent matched filtering
by up to 15 per cent,
depending on where $f_*$ lies relative to the noise
curve minimum.
Coherent, multi-detector searches for continuous waves from
nonaxisymmetric pulsars should be tailored accordingly.

\end{abstract}


\section{Introduction}
\label{sec:acc1}
Nonaxisymmetric
mountains on accreting neutron stars with millisecond spin periods
are promising gravitational wave sources for
long-baseline interferometers like the 
\emph{Laser Interferometer
Gravitational Wave Observatory}
(LIGO).
Such sources
can be detected by coherent matched filtering without
a computationally expensive hierarchical Fourier search
\citep{bra98}, as they
emit continuously at periods and sky positions that are known
a priori from X-ray timing, at least in principle.
Nonaxisymmetric mountains have been invoked to explain why
the spin frequencies $f_*$ of accreting
millisecond pulsars,
measured from X-ray pulses and/or thermonuclear burst oscillations
\citep{cha03,wij03b},
have a distribution that cuts off
sharply above $f_* \approx 0.7$ kHz.
This is well below the centrifugal
break-up frequency for most nuclear equations of state
\citep{coo94},
suggesting that a gravitational wave torque balances the
accretion torque, provided that the stellar ellipticity satisfies
${\epsilon} \sim 10^{-8}$
\citep{bil98}.
Already, the S2 science run on LIGO I
has set upper limits on
${\epsilon}$ for 28 isolated radio pulsars, reaching
as low as
$\epsilon \leq 4.5\times 10^{-6}$ for J2124$-$3358,
following a coherent, multi-detector search synchronized
to radio timing ephemerides
\citep{lig04b}.
Temperature gradients
\citep{bil98,ush00},
large toroidal magnetic fields in the stellar interior
\citep{cut02},
and polar magnetic burial,
in which accreted material accumulates in a polar mountain
confined by the compressed, equatorial
magnetic field
\citep{mel01,pay04,mel05},
have been invoked to account for
ellipticities as large as
${\epsilon}\sim 10^{-8}$.
The latter mechanism is the focus of this paper.

A magnetically confined
mountain is not disrupted by ideal-magnetohydrodynamic
(ideal-MHD)
instabilities, like the Parker instability,
despite the stressed configuration of the magnetic field
\citep{pay05}.
However, magnetospheric disturbances
(driven by accretion rate fluctuations) and
magnetic footpoint motions (driven by stellar tremors)
induce the mountain to
oscillate around its equilibrium position
\citep{mel05}.
In this paper,
we calculate the Fourier spectrum of the
gravitational radiation emitted by the oscillating mountain.
In \S 2, we compute $\epsilon$ as a function of time by
simulating the global oscillation of the mountain numerically
with the ideal-MHD code ZEUS-3D.
In \S 3, we calculate the gravitational wave spectrum
as a function of wave polarization and accreted mass.
The signal-to-noise ratio (SNR)
in the LIGO I and II interferometers is predicted
in \S 4
as a function of $\Ma$,
for situations where the mountain does and does not
oscillate,
and for individual and multiple sources.

\section{Magnetically confined mountain}
\label{sec:burial}

\subsection{Grad-Shafranov equilibria}
\label{sec:gradshafranov}
\begin{figure}
\centering
\plotone{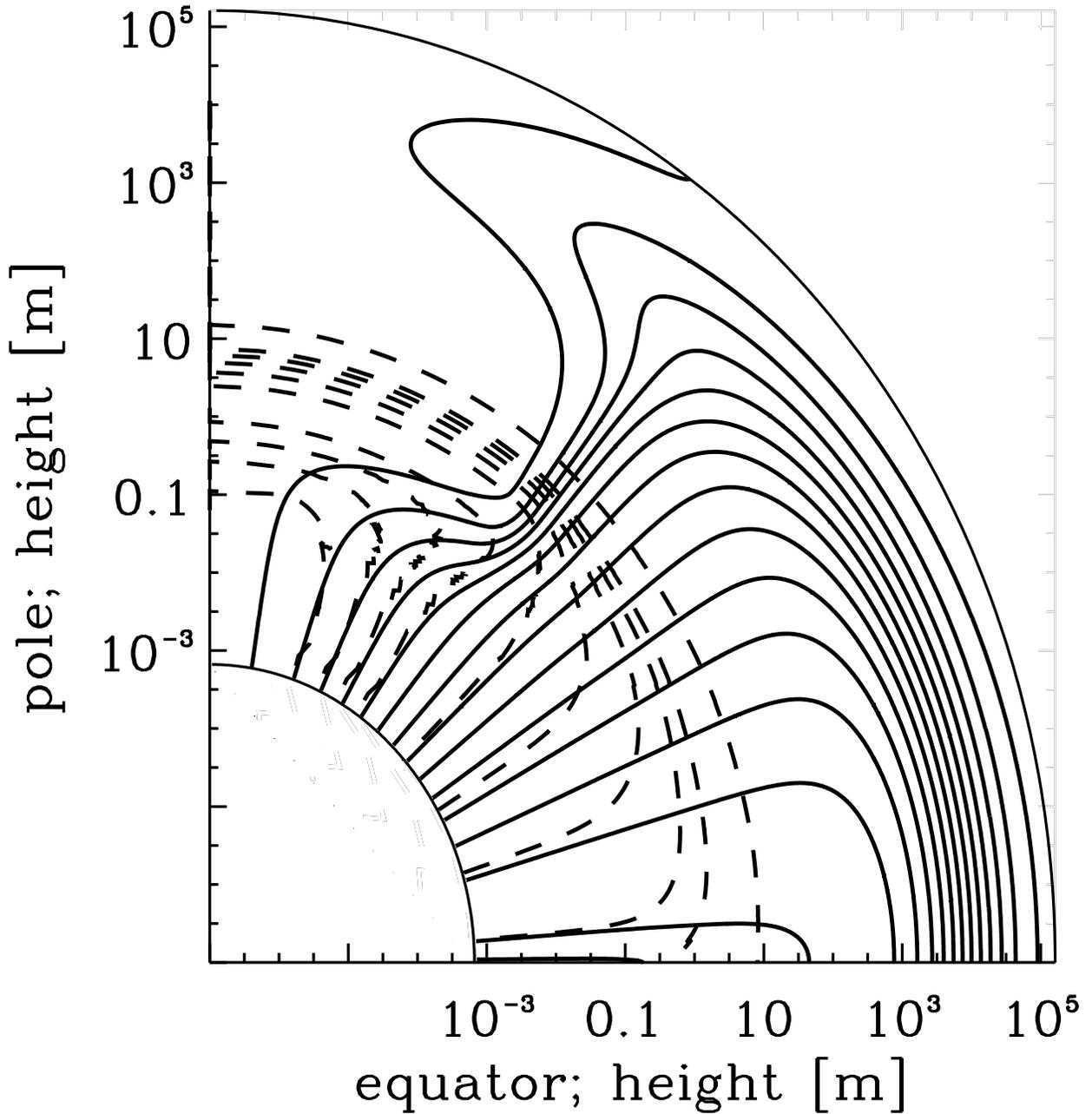}
\caption{\small
Equilibrium magnetic field lines (solid curves)
and density contours (dashed curves) for
$M_{\rm a} = 10^{-5}\Msun$ and $\psi_{\rm a} = 0.1\psi_{*}$.
Altitude is marked on the axes (log scale).
[From \citet{pay04}.]
}
\label{fig:polar}
\end{figure}

During magnetic burial, material accreting onto a neutron star
accumulates in a column at the magnetic polar cap, until
the hydrostatic pressure at the base
of the column overcomes the magnetic tension and
the column spreads equatorward,
compressing the frozen-in magnetic field into
an equatorial magnetic belt or `tutu'
\citep{mel01,pay04}.
Figure \ref{fig:polar} illustrates the equilibrium
achieved for
$\Ma = 10^{-5}\Msun$,
where $\Ma$ is the total accreted mass.
As $\Ma$ increases, the equatorial magnetic belt
is compressed further while
maintaining its overall shape.

In the steady state, the equations of ideal MHD
reduce to the force balance equation
(CGS units)
\begin{equation}
\nabla p + \rho\nabla\Phi - {(4\pi)}^{-1}(\nabla\times {\bf B})\times {\bf B} = 0,
\label{eq:forcebalance}
\end{equation}
where ${\bf B}$, $\rho$,
$p = c_{\rm s}^2\rho$, and
$\Phi(r) = GM_{*}r/R_{*}^{2}$
denote the magnetic field, fluid density,
pressure, and gravitational potential respectively,
$c_{\rm s}$ is the isothermal sound speed,
$M_{*}$ is the mass of the star, and
$R_{*}$ is the stellar radius.
In spherical polar coordinates $(r,\theta,\phi)$,
for an axisymmetric field
${\bf B} = \nabla\psi(r,\theta)/(r\sin\theta)\times\hat{\bf e}_\phi$,
(\ref{eq:forcebalance}) reduces to the Grad-Shafranov equation
\begin{equation} 
\Delta^2\psi = F^{\prime}(\psi)\exp[-(\Phi-\Phi_0)/c_{\rm s}^2],
\label{eq:gradshafranov}
\end{equation}
where $\Delta^{2}$ is the spherical polar
Grad-Shafranov operator,
$F(\psi)$ is an arbitrary function of the magnetic flux $\psi$,
and we set $\Phi_{0} = \Phi(R_{*})$.
In this paper, as in \citet{pay04},
we fix $F(\psi)$ uniquely by connecting 
the initial and final states via the integral form of the flux-freezing
condition, viz.
\begin{equation} 
\frac{dM}{d\psi} = 2\pi\int_C \frac{ds \, \rho}{|\vv{B}|},
\label{eq:fpsi}
\end{equation}
where $C$ is any magnetic field line, and the mass-flux distribution
is chosen to be of the form
$dM/d\psi \propto \exp(-\psi/\psi_{\rm a})$,
where  $\psi_{\rm a}$ is the polar flux,
to mimic magnetospheric accretion
(matter funneled onto the pole).
We also assume north-south symmetry
and adopt the boundary conditions
$\psi =$ dipole
at $r = R_{*}$ (line tying),
$\psi = 0$ at $\theta = 0$,
and $\partial\psi/\partial r = 0$ at large $r$.
Equations (2) and (3)
are solved numerically using an iterative relaxation scheme
and analytically by Green functions,
yielding equilibria like the one in
Figure \ref{fig:polar}.
\clearpage
\begin{figure}
\centering
\plotone{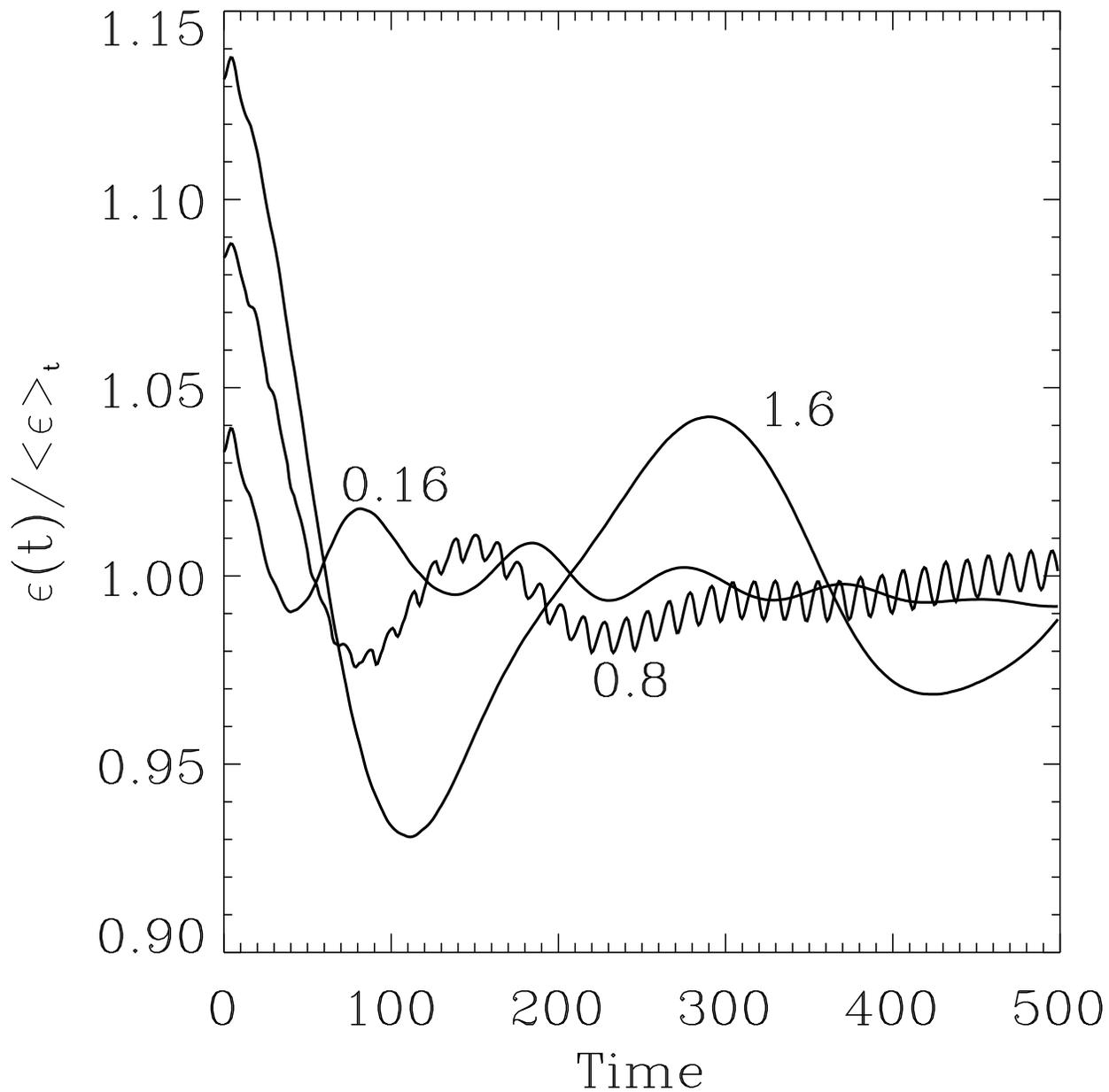}
\caption{\small
Normalized ellipticity $\epsilon(t)/\bar{\epsilon}$ for
$\Ma/\Mc = 0.16, 0.80, 1.6$, with
$\bar{\epsilon} = 8.0\times 10^{-7}, 1.2\times 10^{-6}, 1.3\times 10^{-6}$
respectively for $b = 10$.
Time is measured in units of the Alfv\'en crossing time, $\tau_{\rm A}$.
}
\label{fig:hmean}
\end{figure}
\begin{figure}
\centering
\plottwo{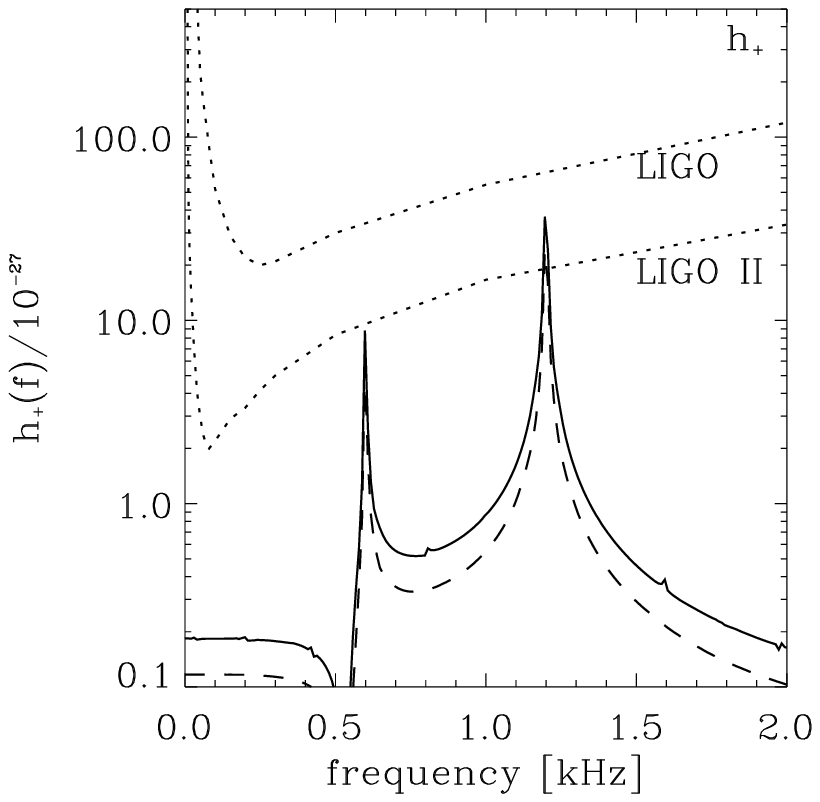}{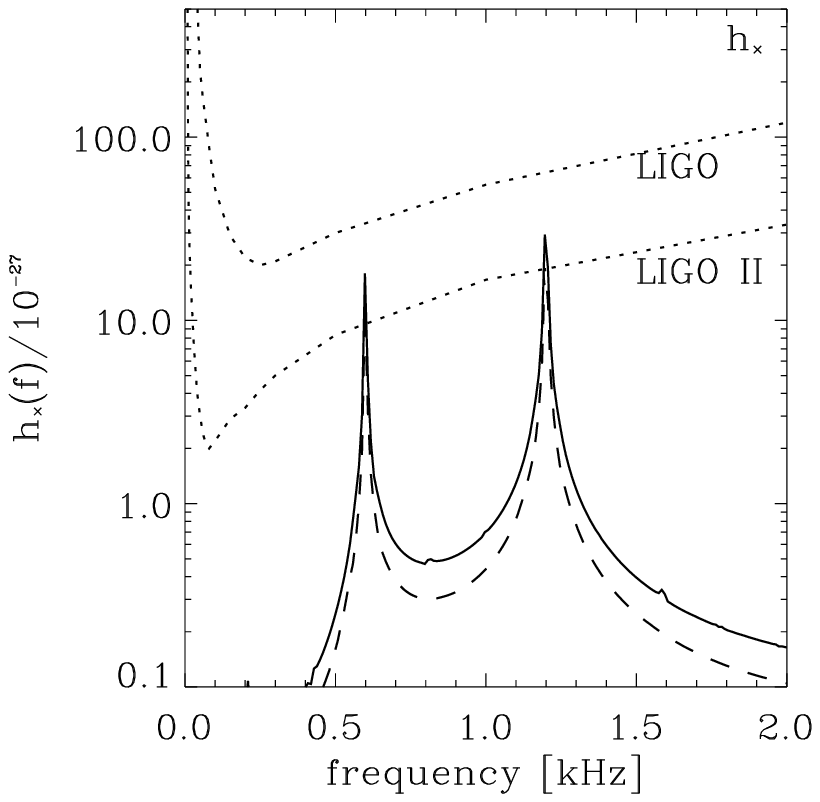}
\plottwo{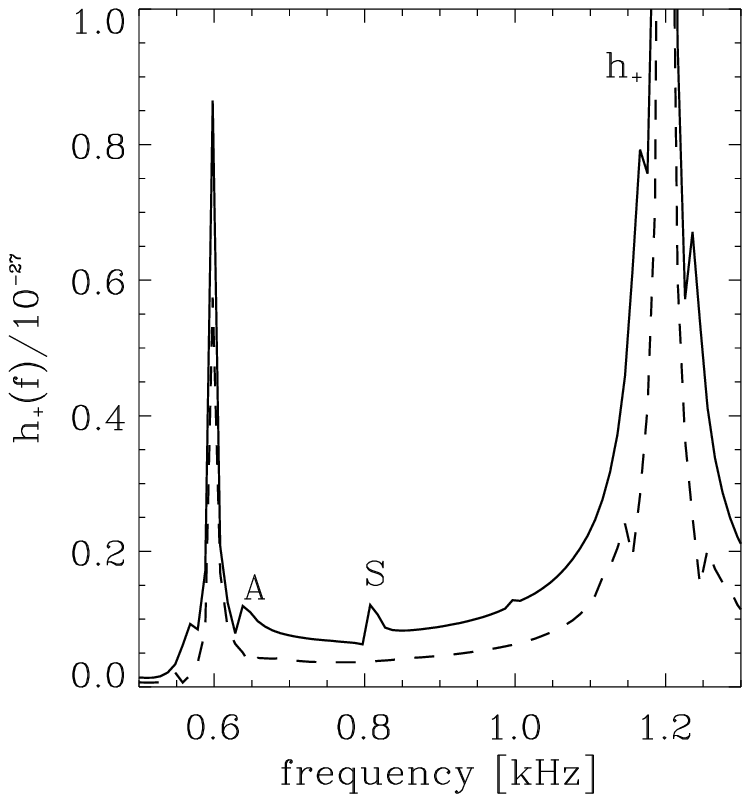}{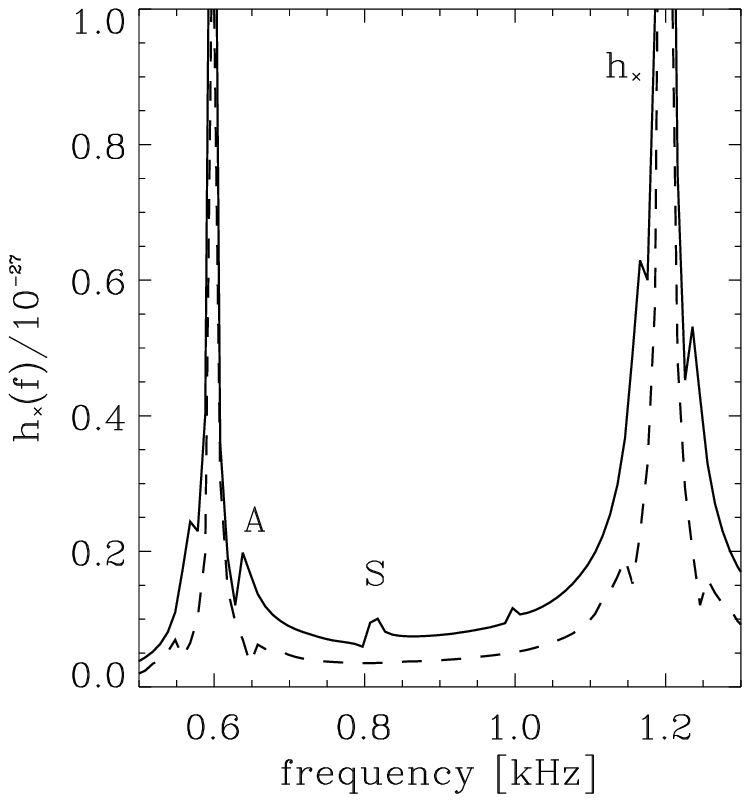}
\caption{\small
(\emph{Top}) Fourier transforms
of the wave strain polarization amplitudes $h_{+}(f)$
(\emph{left})
and $h_{\times}(f)$
(\emph{right})
for
$\Ma/\Mc = 0.16$ (\emph{dashed}) and $0.8$ (\emph{solid}),
compared with the
LIGO I and II noise curves $h_{3/{\rm yr}}$ (see \S \ref{sec:snr})
(\emph{dotted}).
The signals for $\Ma/\Mc = 0.16$ and $0.8$ yield
${\rm SNR} = 2.9$ and $4.4$ respectively 
after $10^{7}$ s.
(\emph{Bottom}).
Zoomed-in view after reducing $h_{+,\times}(f_*)$
and $h_{+,\times}(2f_*)$ artificially
by 90 per cent to bring out the sidebands.
`S' and `A' label the signals induced by  sound- and
Alfv\'en-wave wobbles respectively.
All curves are for
$\alpha = \pi/3$, $i = \pi/3$, $\psi_*/\psi_{\rm a}=10$,
and $d = 10$ kpc.
}
\label{fig:hplus}
\end{figure}
\clearpage
\subsection{Global MHD oscillations}
\label{sec:mhdoscill}
The magnetic mountain is hydromagnetically
stable, even though the confining magnetic field
is heavily distorted.
Numerical simulations using
ZEUS-3D, a multipurpose,
time-dependent, ideal-MHD code for astrophysical fluid dynamics
which uses staggered-mesh
finite differencing and operator splitting
in three dimensions
\citep{sto92},
show that the equilibria from \S \ref{sec:gradshafranov}
are not disrupted by growing Parker or 
interchange modes over a wide range of accreted mass
($10^{-7}\Msun\lesssim\Ma\lesssim 10^{-3}\Msun$)
and intervals as long as
$10^{4}$ Alfv\'en crossing times
\citep{pay05}.

The numerical experiments leading to this conclusion
are performed by loading the output ($\rho$ and $\vv{B}$)
of the Grad-Shafranov code described in \citet{pay04}
into ZEUS-3D, with
the time-step determined by the Courant condition
satisfied by the fast magnetosonic mode.
The code was verified \citep{pay05} by reproducing
the classical Parker
instability of a plane-parallel magnetic field 
\citep{mou74} and the analytic profile of a static, spherical,
isothermal atmosphere.
Coordinates are rescaled in ZEUS-3D to handle the
disparate radial ($c_{\rm s}^2 R_*^2/GM_*$) and latitudinal
($R_*$) length scales.
The stability is confirmed by plotting the
kinetic, gravitational potential, and magnetic energies
as functions of time and observing that the total energy
decreases back to its equilibrium value monotonically 
i.e. the Grad-Shafranov equilibria are
(local) energy minima.
Note that increasing $\rho$ uniformly (e.g. five-fold)
does lead to a
transient Parker instability (localized near the pole) in which
$\lesssim 1 \%$ of the magnetic flux in the `tutu' escapes
through the outer boundary, leaving the
magnetic dipole and mass ellipticity essentially unaffected.

Although the mountain is stable, it does wobble
when perturbed, as sound and
Alfv\'en waves propagate through it
\citep{pay05}.
Consequently, the ellipticity $\epsilon$ of the star
oscillates about its
mean value $\bar{\epsilon}$.
The frequency of the oscillation decreases with $\Ma$,
as described below.
The mean value $\bar{\epsilon}$ increases with $\Ma$
up to a critical mass $\Mc$ and increases with
$\psi_{\rm a}/\psi_{*}$,
as described in \S \ref{sec:gwpolarization}.

In ideal MHD, there is no dissipation
and the oscillations persist for a long time
even if undriven, decaying on the Alfv\'en radiation time-scale
(which is much longer than our longest simulation run).
In reality, the oscillations are also damped by ohmic dissipation,
which is mimicked (imprecisely) by grid-related losses in our work.

To investigate the oscillations quantitatively,
we load slightly perturbed versions of the
Grad-Shafranov equilibria in \S \ref{sec:gradshafranov}
into ZEUS-3D and calculate $\epsilon$ as a function
of time $t$.
Figure \ref{fig:hmean} shows the results of
these numerical experiments.
Grad-Shafranov equilibria are difficult to compute directly
from (\ref{eq:gradshafranov}) and (\ref{eq:fpsi})
for $\Ma \gtrsim 1.6\Mc$,
because the magnetic topology changes and
bubbles form, so instead
we employ a bootstrapping algorithm in ZEUS-3D
\citep{pay05}, whereby
mass is added quasistatically through the
outer boundary and the magnetic field at the outer boundary
is freed to allow
the mountain to equilibrate.
The experiment is performed for 
$r_0/R_{*} = c_{\rm s}^{2} R_{*}/GM_{*} = 2\times 10^{-2}$ 
(to make it computationally tractable) and is then
scaled up to neutron star parameters ($r_0/R_{*} = 5\times 10^{-5}$)
according to $\epsilon\propto (R_{*}/r_{0})^{2}$ and
$\tau_{\rm A}\propto R_*/r_0$,
where $\tau_{\rm A}$ is the Alfv\'en crossing time
over the hydrostatic scale height $r_0$
\citep{pay05}.

The long-period wiggles in Figure \ref{fig:hmean}
represent an Alfv\'en mode with
phase speed $v_{\rm A} \propto \Ma^{-1/2}$;
their period roughly triples from
$100\tau_{\rm A}$ for $\Ma/\Mc = 0.16$ to
$300\tau_{\rm A}$ for $\Ma/\Mc = 1.6$.
Superposed is a shorter-period sound mode, whose
phase speed $c_{\rm s}$ is fixed for all $\Ma$.
Its amplitude is smaller than the Alfv\'en mode;
it appears in all three curves in Figure \ref{fig:hmean}
as a series of small kinks for $t \lesssim 50\tau_{\rm A}$,
and is plainly seen at all $t$ for $\Ma/\Mc = 0.8$.
As $\Ma$ increases,
the amplitude of the Alfv\'en component at
frequency $f_{\rm A} \sim 17 (\Ma/\Mc)^{-1/2}$ Hz
is enhanced.  By contrast,
the sound mode stays fixed at a frequency $f_{\rm S}\sim 0.4$ kHz,
while its amplitude peaks at $\Ma\sim\Mc$
\citep{pay05}.

\section{Frequency spectrum of the gravitational radiation}
\label{sec:gwfreq}
In this section, we predict the frequency spectrum of the
gravitational-wave signal
emitted by
freely oscillating 
and stochastically perturbed magnetic mountains
in the standard orthogonal polarizations.

\subsection{Polarization amplitudes}
\label{sec:gwpolarization}
The metric perturbation
for a biaxial rotator can be written in
the transverse-traceless gauge as
$h_{ij}^{\rm TT} = h_+ \, e_{ij}^+ \ + \ h_\times \, e_{ij}^\times$, 
where
$e_{ij}^+$ and $e_{ij}^\times$ are the
basis tensors for the $+$ and $\times$ polarizations
and the wave strains $h_+$ and $h_{\times}$ are given by
\citep{zim79,bon96}
\begin{eqnarray}
\label{eq:hplus}
   h_+ & = & h_0 \sin\alpha [
        \cos\alpha\sin i\cos i \cos(\Omega t) \nonumber \\
      & & \qquad \qquad
 - \sin\alpha ({1+\cos^2 i}) \cos(2\Omega t) ] \label{e:h+,gen} \, , \\
 \label{eq:hcross}
    h_\times & = & h_0 \sin\alpha [
        \cos\alpha\sin i\sin(\Omega t) \nonumber \\
	& & \qquad \qquad
 - 2 \sin\alpha \cos i \sin(2\Omega t) ] \ ,
\end{eqnarray}
with\footnote{
Our $h_{0}$ is half that given by Eq. (22) of \citet{bon96}.
}
\begin{equation} \label{eq:h0}
        h_0 = {2 G} {I}_{zz}{\epsilon}
	       {\Omega^2/d c^4} \, .
\end{equation}
Here,
$\Omega = 2\pi f_*$ is the stellar angular velocity,
$i$ is the angle between the
rotation axis $\vv{e}_{z}$ and the 
line of sight,
$\alpha$ is the angle between $\vv{e}_{z}$ and the
magnetic axis of symmetry,
and $d$ is the distance of the source from Earth.

The ellipticity is given by
${\epsilon} = |I_{zz}-I_{yy}|/I_0$,
where
$I_{ij}$ denotes the moment-of-inertia tensor and
$I_0 = \frac{2}{5} M_* R_*^2$.
In general, $\epsilon$ is a function of $t$;
it oscillates about a mean value $\bar{\epsilon}$,
as in Figure \ref{fig:hmean}.
Interestingly, the oscillation frequency can approach $\Omega$
for certain values of $\Ma$
(see \S \ref{sec:freeoscill}),
affecting the detectability of the source
and complicating the design of matched filters.
The mean ellipticity is given by
\citep{mel05}
\begin{equation}
\bar{\epsilon} =
\begin{cases}
1.4 \times 10^{-6} \
\left(\frac{\Ma}{10^{-2}\Mc}\right)\left(\frac{B_{*}}{10^{12} {\rm \, G}}\right)^{2} \quad  \Ma\ll\Mc \\
\frac{5\Ma}{2M_{*}}\left(1 - \frac{3}{2b}\right)\left(1+\frac{\Ma b^2}{8\Mc}\right)^{-1} \quad\quad\,\,\,\, \Ma\gtrsim\Mc
\label{eq:epsilonbig}
\end{cases}
\end{equation}
where
$\Mc = G M_{*} B_{*}^{2} R_{*}^{2}/(8 c_{\rm s}^{4})$
is the critical mass beyond which the accreted matter
bends the field lines appreciably,
$b = \psi_{*}/\psi_{\rm a}$ is the hemispheric to polar flux ratio,
and $B_{*}$ is the polar magnetic field strength prior to accretion.
For
$R_{*} = 10^6$ cm, $c_{\rm s}=10^8$ cm s$^{-1}$, and
$B_{*} = 10^{12}$ G, we find
$\Mc = 1.2\times 10^{-4}\Msun$.
The maximum ellipticity,
$\bar{\epsilon}\rightarrow 20\Mc/(M_* b^2)
\sim 10^{-5}(b/10)^{-2}$ as
$\Ma\rightarrow\infty$,
greatly exceeds previous estimates,
e.g.
$\bar{\epsilon}\sim 10^{-12}$
for an undistorted dipole
\citep{kat89,bon96}, 
due to the heightened
Maxwell stress exerted by the compressed
equatorial magnetic belt.
Note that $\epsilon(t)$ is computed using ZEUS-3D for
$b = 3$ (to minimize numerical errors)
and scaled to larger $b$ using
(\ref{eq:epsilonbig}).

\subsection{Natural oscillations}
\label{sec:freeoscill}
We begin by studying the undamped, undriven oscillations of
the magnetic mountain when it is ``plucked",
e.g. when a
perturbation is introduced via
numerical errors when the equilibrium is
translated from the Grad-Shafranov grid to the
ZEUS-3D grid
\citep{pay05}.
We calculate
$h_{\times}(t)$ and $h_{+}(t)$ for $f_* = 0.6$ kHz
from (\ref{eq:hplus}) and (\ref{eq:hcross})
and display the Fourier transforms
$h_{\times}(f)$ and $h_{+}(f)$ in Figure
\ref{fig:hplus}
for two values of $\Ma$.
The lower two panels provide an enlarged view
of the spectrum around the peaks;
the amplitudes at $f_{*}$ and $2 f_{*}$ are
divided by ten to help bring out
the sidebands.

In the enlarged panels, we see that
the principal carrier frequencies
$f=f_*, \, 2f_*$ are flanked by two
lower frequency peaks arising from
the Alfv\'en mode of the oscillating mountain
(the rightmost of which is labelled `A').
Also, there is a peak (labelled `S') displaced by
$\Delta f \sim 0.4$ kHz from the principal
carriers which arises from the
sound mode; it is clearly visible 
for $\Ma/\Mc = 0.8$
and present, albeit imperceptible without magnification,
for $\Ma/\Mc = 0.16$.
Moreover,
$\epsilon$ diminishes gradually over many $\tau_{\rm A}$
(e.g. in Figure \ref{fig:hmean}, for $\Ma/\Mc = 0.16$,
$\epsilon$ drifts from $1.02\bar{\epsilon}$ to
$0.99\bar{\epsilon}$
over $500\tau_{\rm A}$),
causing the peaks at $f = f_*,\, 2f_*$ to broaden.
As $\Ma$ increases, this broadening increases;
the frequency of the Alfv\'en component scales as
$f_{\rm A} \propto \Ma^{-1/2}$ and
its amplitude increases $\propto\Ma^{1/2}$
(see \S \ref{sec:mhdoscill}); and
the frequency of the sound mode stays fixed at $f_{\rm S}\sim 0.4$ kHz
\citep{pay05}.
Note that
these frequencies must be scaled to convert
from the numerical model ($r_0/R_{*} = 2\times 10^{-2}$) to
a realistic star ($r_0/R_{*} = 5\times 10^{-5}$);
it takes proportionally longer for the signal
to cross the mountain \citep{pay05}.

\subsection{Stochastically driven oscillations}
\begin{figure}
\centering
\plottwo{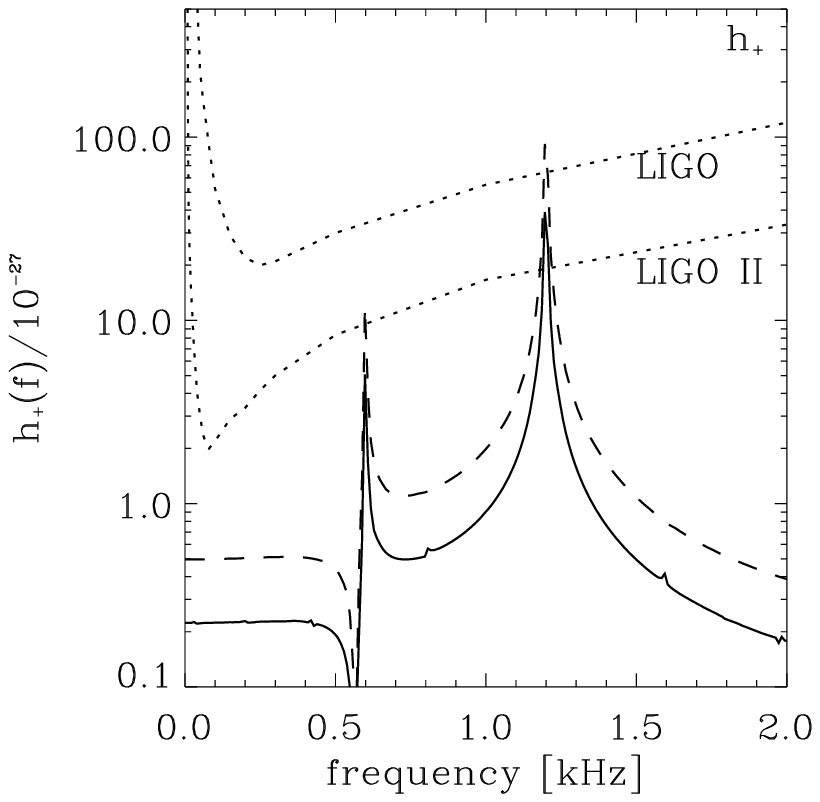}{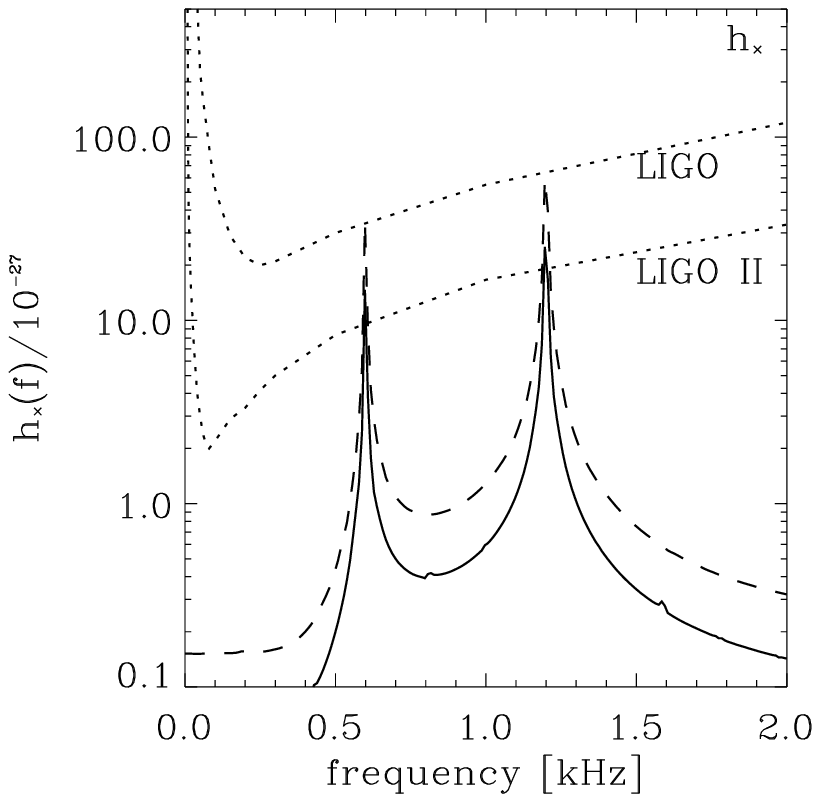}
\plottwo{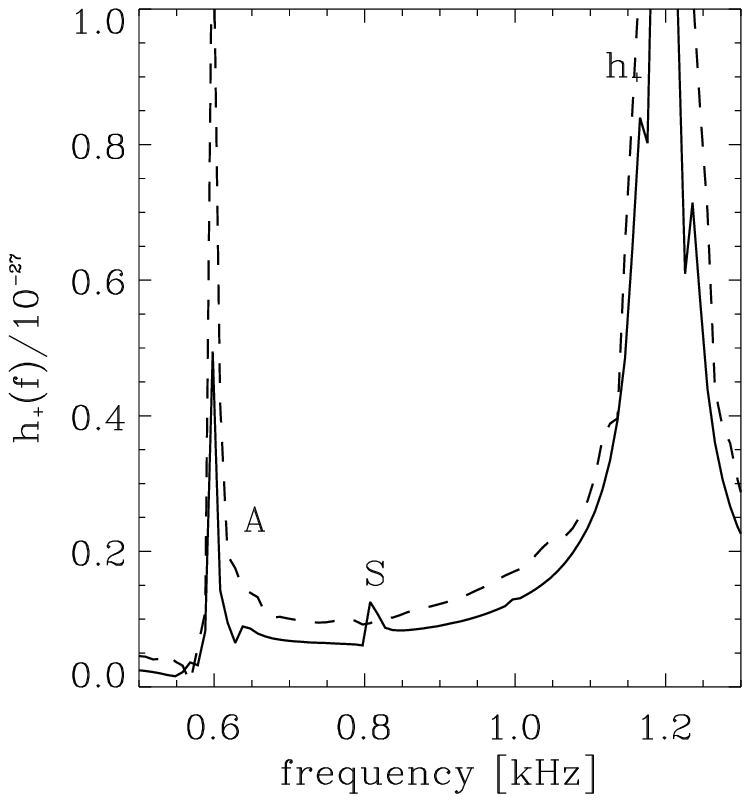}{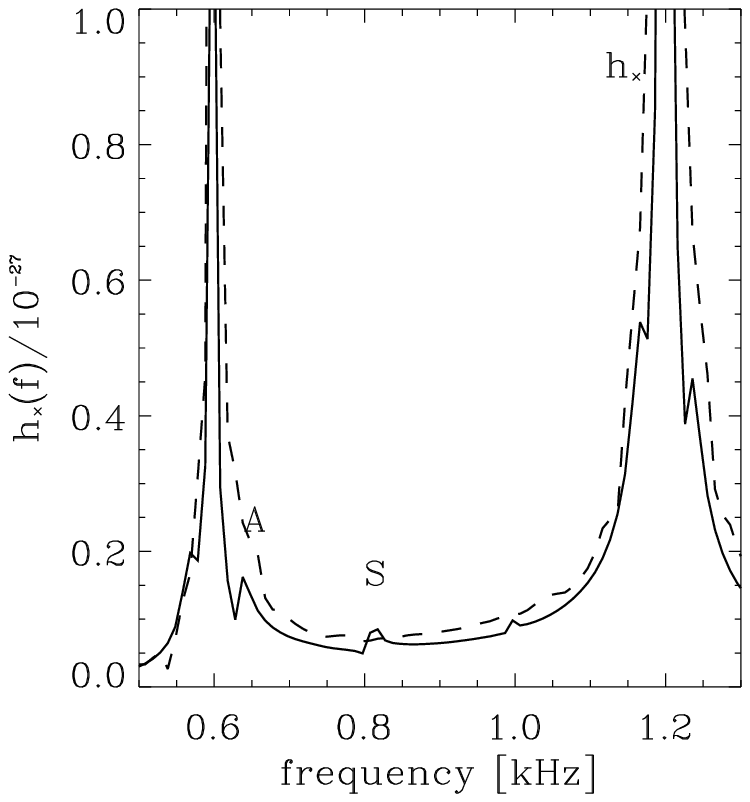}
\caption{\small
(\emph{Top}) Fourier transforms
of the wave strain polarization amplitudes $h_{+}(f)$
(\emph{left})
and $h_{\times}(f)$
(\emph{right})
for
$\Ma/\Mc = 0.8$ with stochastic (\emph{dashed}) and natural (\emph{solid}),
oscillations.
compared with the
LIGO I and II noise curves $h_{3/{\rm yr}}$ (see \S \ref{sec:snr})
(\emph{dotted})
corresponding to 99\% confidence after $10^{7}$ s.
(\emph{Bottom})
Zoomed in view with $h_{+,\times}(f_*)$ and $h_{+,\times}(2f_*)$
artificially reduced
by 90 per cent to bring out the sidebands.
`S' and `A' label the signal induced by sound- and Alfv\'en-wave
wobbles respectively.
All curves are for
$\alpha = \pi/3$, $i = \pi/3$, $\psi_{*}/\psi_{\rm a} = 10$,
and $d = 10$ kpc.
}
\label{fig:stochastic}
\end{figure}

We now study the response of the mountain to a more complex
initial perturbation.
In reality,
oscillations may be excited stochastically by incoming blobs of accreted
matter
\citep{wyn95} or starquakes that perturb the magnetic footpoints
\citep{lin98}.
To test this, we perturb the Grad-Shafranov equilibrium
$\psi_{\rm GS}$
with a truncated series of spatial modes such that
\begin{equation}
\psi = \psi_{\rm GS}\{1 + \Sigma_{n}\delta_{n}\sin[m\pi(r-R_{*})/(r_{\rm max}-R_{*})]\sin(m\theta)\}
\end{equation}
at $t = 0$,
with mode amplitudes scaling according to a power law
$\delta_{n} = 0.25m^{-1}$,
$m = 2n+1$, $0\leq n\leq 3$, 
as one might expect for a noisy process.
We place
the outer grid boundary at
$r_{\rm max} = R_* + 10 r_0$.
Figure \ref{fig:stochastic} compares the resulting spectrum
to that of the free
oscillations in \S \ref{sec:freeoscill} for $\Ma/\Mc = 0.8$.
The stochastic oscillations increase the overall
signal strength at and away from
the carrier frequencies $f_*$ and $2f_*$.
The emitted power also
spreads further in frequency,
with the full-width half-maximum of the principal carrier peaks
measuring $\Delta f \approx 0.25$ kHz
(c.f. $\Delta f \approx 0.2$ kHz in Figure \ref{fig:hplus}).
However, the overall shape of the spectrum remains unchanged.
The Alfv\'en and sound peaks are partially washed out by the
stochastic noise
but remain perceptible upon magnification.
The signal remains above the LIGO II noise
curves in Figure \ref{fig:stochastic};
in fact, its detectability can (surprisingly) be
enhanced, as we show below in \S \ref{sec:snr}.
\clearpage
\clearpage
\section{Signal-to-noise ratio}
\label{sec:snr}
In this section,
we investigate how oscillations of the mountain affect
the SNR of such sources, and how the SNR varies with
$\Ma$.
In doing so, we generalize expressions for the SNR and
characteristic wave strain $h_{\rm c}$ in the literature
to apply to nonaxisymmetric neutron stars oriented
with arbitrary $\alpha$ and $i$.

\subsection{Individual versus multiple sources}
The signal received at Earth from an individual source
can be written as
$h(t) = F_{+}(t)h_{+}(t) + F_{\times}(t)h_{\times}(t)$,
where $F_{+}$ and $F_{\times}$ are detector beam-pattern functions
($0\leq|F_{+,\times}|\leq 1$)
which depend on the sky position of the source as
well as $\alpha$ and $i$ \citep{tho87}.
The squared SNR is then \citep{cre03}\footnote{
This is twice the SNR defined in Eq. (29) of \citet{tho87}.}
\begin{equation}
\label{eq:snr}
\frac{S^2}{N^2} = 4\int_{0}^{\infty}df\, \frac{|h(f)|^2}{S_{h}(f)} \, ,
\end{equation}
where
$S_{h}(f) = |h_{3/{\rm yr}}(f)|^2$ is the one-sided spectral
density sensitivity function of the detector
(Figures \ref{fig:hplus} and \ref{fig:stochastic}),
corresponding to the
weakest source detectable with 99 per cent confidence in 
$10^{7}$ s of integration time, if the frequency
and phase of the signal
at the detector are known in advance
\citep{bra98}.

A characteristic amplitude $h_{\rm c}$
and frequency $f_{\rm c}$ can also be defined
in the context of periodic sources.
For an individual source, where we know $\alpha$, $i$,
$F_+$ and $F_{\times}$ in principle, the definitions 
take the form
\begin{equation}
\label{eq:fc}
f_{\rm c} = \left[\int_{0}^{\infty}df\, \frac{|h(f)|^2}{S_h(f)}\right]^{-1} \
\left[\int_{0}^{\infty} df\, f \frac{|h(f)|^2}{S_h(f)}\right]\, ,
\end{equation}
and
\begin{equation}
\label{eq:hc}
h_{\rm c} =\frac{S}{N}[S_h(f_{\rm c})]^{1/2} \, .
\end{equation}
These definitions are valid not only
in the special case of an individual source with
$\alpha = \pi/2$
(emission at $2f_*$ only) but also more generally
for arbitrary $\alpha$
(emission at $f_*$ and $2f_*$).
Using (\ref{eq:hplus}), (\ref{eq:hcross}), (\ref{eq:fc}) and (\ref{eq:hc}),
and assuming for the moment that $\epsilon$ is constant
(i.e. the mountain does not oscillate), we obtain
\begin{equation}
f_{\rm c} = f_*(\chi A_1+2 A_2)/(\chi A_1+A_2) \, ,
\end{equation}
\begin{equation}
\frac{S}{N} = 
h_{0}[S_{h}(2f_*)]^{-1/2}(\chi A_1+A_2)^{1/2}\sin\alpha
\end{equation}
with
$A_1 = \cos^2\alpha\sin^2 i(F_+\cos i +F_{\times})^2$,
$A_2 = \sin^2\alpha[F_+(1+\cos^2 i)+2F_{\times}\cos i]^2$,
$\chi = S_h(2f_*)/S_h(f_*)$,
and
$\eta = S_h(f_c)/S_h(f_*)$.
In the frequency range $0.2\leq f \leq 3$ kHz,
the LIGO II noise curve is fitted well by
$h_{3/{\rm yr}}(f) = 10^{-26}(f/0.6 {\rm kHz})$ Hz$^{-1/2}$
\citep{bra98}, implying $\chi = 4$.
As an example, for
($\alpha,i)=(\pi/3,\pi/3)$, we obtain
$f_{\rm c} = 1.67f_*$,
$h_{\rm c} = 1.22h_0$
and
$S/N
= 2.78(f_*/0.6{\rm kHz})(\epsilon/10^{-6})(d/10 {\rm kpc})^{-1}$.
In the absence of specific knowledge of the source position,
we take
$F_{\times}=F_+ = 1/\sqrt{5}$
(for motivation, see below).

If the sky position and orientation of
individual sources are unknown, it is sometimes useful
to calculate the orientation- and polarization-averaged
amplitude $\bar{h}_{\rm c}$ and frequency
$\bar{f}_{\rm c}$.
To do this, one cannot assume $\alpha = \pi/2$,
as many authors do
\citep{tho87,bil98,bra98};
sources in an ensemble generally emit at $f_*$ and $2f_*$.
Instead, we replace $|h(f)|^2$ by
$\langle |h(f)|^2\rangle$
in (\ref{eq:snr}), (\ref{eq:fc}) and (\ref{eq:hc}),
defining the average as
$\langle Q\rangle = \int_{0}^{1}\int_{0}^{1}Q \, d(\cos\alpha) \, d(\cos i)$.
This definition
is not biased towards sources with small $\alpha$;
we prefer it to the average
$\langle Q\rangle_{2} = \pi^{-1}\int_{0}^{1}\int_{0}^{\pi}Q \, d\alpha \, d(\cos i)$,
introduced in Eq. (87) of \citet{jar98}.
Therefore, given an ensemble of
neutron stars with mountains
which are not oscillating, we take
$\langle F_{+}^2\rangle = \langle F_{\times}^2\rangle = 1/5$ and
$\langle F_+ F_{\times}\rangle = 0$
[Eq. (110) of \citet{tho87},
c.f. \citet{bon96,jar98}],
average over $\alpha$ and $i$ to get
$\langle A_1\sin^2\alpha\rangle = 8/75$ and
$\langle A_2\sin^2\alpha\rangle = 128/75$, and hence arrive at
$\bar{f}_{\rm c} = 1.80f_*$,
$\bar{h}_{\rm c} = 1.31 h_0$
and
$\langle S^2/N^2\rangle^{1/2} =
2.78(f_*/0.6{\rm kHz})(\epsilon/10^{-6})(d/10 {\rm kpc})^{-1}$.
This ensemble-averaged SNR is similar to the
non-averaged value for $(\alpha, i) = (\pi/3,\pi/3)$, 
a coincidence of the particular choice.

Our predicted SNR, averaged rigorously
over $\alpha$ and $i$ as above, is $(2/3)^{1/2}$
times smaller than it would be for $\alpha = \pi/2$,
because the (real) extra power at
$f_*$ does not make up for the (artificial)
extra power that comes from
assuming that all sources
are maximal ($\alpha = \pi/2$) emitters.
Our value of $\bar{h}_{\rm c}$
is $9/10$ of the value
of $h_{\rm c}$
quoted widely in the literature
\citep{tho87,bil98,bra98}.
The latter authors, among others, assume $\alpha = \pi/2$
and average over $i$, whereas we
average over $\alpha$ and $i$ to account for signals
at both $f_*$ and $2f_*$;
they follow Eq. (55) of \citet{tho87},
who, in the context of \emph{bursting}
rather than continuous-wave sources,
multiplies $h_{\rm c}$ by $(2/3)^{1/2}$
to reflect a statistical preference
for sources with directions and polarizations that give larger
SNRs (because they can be seen out to greater distances);
and they assume $f_{\rm c} = 2f_*$ instead of
$f_{\rm c} = 9f_*/5$ as required by (\ref{eq:fc}).

\subsection{Oscillations versus static mountain}
We now compare a star with an oscillating
mountain against a star whose mountain is in equilibrium.
We compute (\ref{eq:fc}) and (\ref{eq:hc}) directly
from $\epsilon(t)$ as generated by ZEUS-3D
(see \S \ref{sec:burial} and \ref{sec:gwpolarization}),
i.e. without assuming that
$h_{+}(f)$ and $h_{\times}(f)$ are pure $\delta$ functions
at $f = f_*, \, 2f_*$.

Table \ref{table:snr} lists the SNR and
associated characteristic quantities for
three $\Ma$ values (and $b = 10$)
for both the static and oscillating mountains.
The case of a particular $\alpha$ and $i$
($\alpha = i = \pi/3$) is shown along with the
average over $\alpha$ and $i$
\citep{tho87,bil98,bra98}.
We see that the oscillations increase the SNR by up to
$\sim 15$ per cent;
the peaks at $f= f_*, \, 2f_*$ are the same amplitude as for a
static mountain, but
additional signal is contained in the sidebands.
At least one peak exceeds the LIGO II noise curve in
Figure \ref{fig:hplus}
in each polarization.

\subsection{Detectability versus $\Ma$}
The SNR increases with $\Ma$, primarily because $\bar{\epsilon}$
increases.
The effect of the oscillations is more complicated:
although
the Alfv\'en sidebands increase in amplitude as $\Ma$ increases,
their frequency displacement from $f = f_*$ and $f = 2f_*$
decreases,
as discussed in \S \ref{sec:freeoscill},
so that the extra power is confined in a narrower
range of $f$.  However,
$\epsilon$ and hence the SNR plateau when
$\Ma$ increases above $\Mc$
(see \S \ref{sec:gwpolarization}).
The net result is that
increasing $\Ma$ by a factor of 10 raises the SNR
by less than a factor of two.
The SNR saturates at $\sim 3.5$ when averaged
over $\alpha$ and $i$ (multiple sources),
but can reach $\sim 6$ for a particular source
whose orientation is favorable.
For our parameters, an accreting neutron star
typically
becomes detectable with LIGO II once it has
accreted $\Ma \gtrsim 0.1\Mc$.
The base of the mountain may be at a depth
where the ions are crystallized, but
an analysis of the crystallization properties
is beyond the scope of this paper.
\clearpage
\begin{table}
\begin{center}
\caption{
Signal-to-noise ratio
}
\begin{tabular}{ccccc}
\hline
\hline
$f_*$ [kHz] & $\Ma/10^{-4}\Msun$ & $f_{\rm c} [{\rm kHz}]$   & $h_{\rm c}/10^{-25}$ & SNR \\
\hline
 & \quad\quad Static &$\alpha=\pi/3$ &$i=\pi/3$ & \\
\hline
0.6         & $0.16$ & 1.003 & 0.83 & 2.22 \\
0.6         & $0.8$   & 1.003 & 1.24   & 3.34 \\
0.6         & $1.6$ & 1.003 & 1.35   & 3.61 \\
\hline
& \quad\quad Static & $\langle \, \rangle_{\alpha}$ & $\langle \, \rangle_{i}\quad\quad $ & \\
\hline
0.6         & $0.16$ & 1.08 & 0.89 & 2.22 \\
0.6         & $0.8$   & 1.08 & 1.33  & 3.34 \\
0.6         & $1.6$ & 1.08 & 1.44  & 3.61 \\
\hline
 & Oscillating &$\alpha=\pi/3$ &$i=\pi/3$ & \\
\hline
0.6         & $0.16$ & 1.008 & 1.40    & 2.63 \\
0.6         & $0.8$   & 1.003 & 2.15   & 4.02 \\
0.6         & $1.6$ & 1.004 & 2.27    & 4.25 \\
\hline
 & Oscillating & $\langle \, \rangle_{\alpha}$ & $\langle \, \rangle_{i}\quad\quad $ & \\
\hline
0.6         & $0.16$ & 1.056 & 1.40 & 2.45 \\
0.6         & $0.8$   & 1.048 & 2.14  & 3.74 \\
0.6         & $1.6$ & 1.048 & 2.26  & 3.95 \\
\hline
\end{tabular}
\label{table:snr}
\end{center}
\end{table}
\clearpage

\section{DISCUSSION
 \label{sec:acc4}}
A magnetically confined mountain forms at the magnetic poles of an accreting
neutron star during the process of magnetic burial.
The mountain, which is generally offset from the spin axis,
generates gravitational waves at $f_*$ and $2 f_*$.
Sidebands in the gravitational-wave spectrum appear around
$f_*$ and $2f_*$ due to
global MHD oscillations of the mountain
which may be excited by stochastic variations in
accretion rate (e.g. disk instability) or
magnetic footpoint motions (e.g. starquake).
The spectral peaks at
$f_*$ and $2f_*$ are
broadened, with full-widths half-maximum
$\Delta f \approx 0.2$ kHz.
We find that the SNR increases
as a result of these oscillations by up to 15 per cent
due to additional signal from around the peaks.

Our results suggest that
sources such as
SAX J1808.4$-$3658
may be detectable by next generation long-baseline
interferometers like LIGO II.
Note that,
for a neutron star accreting matter at the rate
$\dot{M}_{\rm a} \approx 10^{-11} \Msun \, {\rm yr}^{-1}$
(like SAX J1808.4$-$3658),
it takes only $10^{7}$ yr to reach
$S/N > 3$.\footnote{ On the other hand,
EOS 0748$-$676, whose accretion rate is
estimated to be at least ten times greater,
at $\dot{M}_{\rm a} \gtrsim 10^{10}\Msun {\rm yr}^{-1}$,
has $f_* = 45$ Hz (from burst oscillations) and does not pulsate,
perhaps because hydromagnetic spreading has already proceeded further
($\mu \lesssim 5\times 10^{27} {\rm G \, cm}^{-3}$ \citep{vil04}.
}
The characteristic wave strain
$h_{\rm c} \sim 4\times 10^{-25}$ is also comparable to that
invoked by \citet{bil98} to explain the observed range of
$f_*$ in low-mass X-ray binaries.
An observationally testable scaling between
$h_{\rm c}$ and the magnetic dipole moment {\boldmath $|\mu|$}
has been predicted
\citep{mel05}.

The analysis in \S \ref{sec:gwfreq} and \S \ref{sec:snr} applies
to a biaxial star whose principal axis of inertia coincides
with the magnetic axis of symmetry and
is therefore inclined with respect to the angular momentum axis
{\boldmath $J$}
in general (for $\alpha \neq 0$).
Such a star precesses \citep{cut01}, a fact neglected in our analysis
up to this point in order to
maintain consistency with \citet{bon96}.
The latter authors explicitly disregarded precession,
arguing that most of the stellar interior
is a fluid (crystalline crust $\lesssim 0.02 M_*$),
so that the precession frequency is reduced by
$\sim 10^{5}$ relative to a rigid star \citep{pin74}.
Equations (\ref{eq:hplus}) and (\ref{eq:hcross}) display
this clearly.
They are structurally identical to the equations in both
\citet{bon96} and \citet{zim79}, but these papers solve
different physical problems. 
In \citet{zim79}, $\Omega$ differs from the pulsar spin frequency
by the body-frame precession frequency,
as expected for a precessing, rigid, Newtonian star,
whereas in \citet{bon96}, $\Omega$
exactly equals the pulsar spin frequency,
as expected for a (magnetically) distorted (but nonprecessing)
fluid star.
Moreover, $\theta$ (which replaces $\alpha$) in \citet{zim79}
is the angle between the angular momentum vector {\boldmath $J$}
(fixed in inertial space) and the
principal axis of inertia $\vv{e}_3$, whereas $\alpha$ in \citet{bon96}
is the angle between
the rotation axis {\boldmath $\Omega$} and axis of symmetry
{\boldmath $\mu$} of the (magnetic) distortion.
Both interpretations match on time-scales that are short compared
to the free precession time-scale
$\tau_{\rm p} \approx (f_*\epsilon)^{-1}$,
but the quadrupole moments computed in this paper
($\epsilon \sim 10^{-7}$) and invoked by \citet{bil98} to explain
the spin frequencies of low-mass X-ray binaries
($10^{-8}\leq\epsilon\leq 10^{-7}$) predict
$\tau_{\rm p}$ of order hours to days.
The effect is therefore likely to be observable,
unless internal damping proceeds rapidly.
Best estimates \citep{jon02} of the dissipation time-scale give
$\approx 3.2 {\rm \, yr \, }(Q/10^4)(0.1 {\rm kHz}/f_*)$
$(I_0/10^{44} {\rm g \, cm}^{2})$
$(10^{38} {\rm g \, cm}^{2}/I_{\rm d})$,
where
$I_{\rm d}$ is the piece of the moment of inertia
that ``follows" $\vv{e}_{3}$ (not {\boldmath $\Omega$}),
and $400\lesssim Q \lesssim 10^{4}$
is the quality factor of the internal damping
[e.g. from electrons scattering off superfluid vortices
\citep{alp88}].\footnote{
Precession has been detected in the isolated radio pulsar
PSR B1828$-$11
\citep{sta00,lin01}.
Ambiguous evidence also exists for long-period ($\sim$ days)
precession in the Crab \citep{lyn88}, Vela \citep{des96},
and PSR B1642$-$03 \citep{sha01}.
Of greater relevance here, it may be that
Her~X-1 precesses \citep[e.g.][]{sha98}.  This object is
an accreting neutron star whose precession may be
continuously driven.
}
\clearpage
\begin{table}
\begin{center}
\caption{
Precession scenarios and associated gravitational wave signals}
\begin{tabular}{cccc}
\hline
\hline
        & biaxial, $\vv{e}_3 \|$ {\boldmath $\Omega$} & triaxial, $\vv{e}_3 \|$ {\boldmath $\Omega$} & $\vv{e}_3 \nparallel$ {\boldmath $\Omega$} \\
	\hline
	$\vv{e}_3 \|$ {\boldmath $\mu$} &  zero GW & GW at $2f_*$ & GW near $f_*$ and $2f_*$ \\ 
        & no precession        & no precession & precession             \\
	& no pulses            & no pulses     & pulses        \\
	\hline
	$\vv{e}_3 \nparallel$ {\boldmath $\mu$} & zero GW & GW at $2f_*$  & GW near $f_*$ and $2f_*$  \\
	&  no precession       & no precession & precession             \\
	&  pulses              & pulses        & pulses \\
\hline
\hline
\end{tabular}
\tablecomments{Here,
$\vv{e}_3$ is the principal axis of inertia,
{\boldmath $\mu$} is the axis of the magnetic dipole,
{\boldmath $\Omega$} is the spin axis, and
$f_*$ is the spin frequency.
Entries containing
$f_*$ and/or $2f_*$ indicate gravitational wave
emission at (or near, in the case of precession) those frequencies;
entries labelled `zero GW' indicate no
gravitational wave emission.
We also specify whether or not each scenario
admits X-ray pulsations.
}
\label{table:pulsargw}
\end{center}
\end{table}
\clearpage
Some possible precession scenarios are summarized in Table \ref{table:pulsargw}.
If we attribute persistent X-ray pulsations to magnetic funnelling
onto a polar hot spot, or to a magnetically anisotropic
atmospheric opacity, then the angle between
{\boldmath $\mu$} and {\boldmath $\Omega$} must be large,
leading to precession with a large wobble angle, which would
presumably be damped on short time-scales unless it is
driven
(cf. Chandler wobble).
Such a pulsar emits gravitational waves at a frequency near
$f_*$ (offset by the body-frame precession frequency) and $2f_*$.
However, 
the relative orientations of {\boldmath $\mu$},
{\boldmath $\Omega$}, and $\vv{e}_{3}$ are determined when the
crust of the newly born neutron star crystallizes after birth
and subsequently by accretion torques.
This is discussed in detail by \citet{mel00a}.
If viscous dissipation in the fluid star forces
{\boldmath $\Omega$} to align with {\boldmath $\mu$}
before crystallization,
and if the symmetry axis of the crust when it crystallizes
is along {\boldmath $\Omega$},
then $\vv{e}_3$
(of the crystalline crust plus the subsequently accreted mountain),
{\boldmath $\mu$}, and {\boldmath $\Omega$} are
all parallel and there is no precession
(nor, indeed, pulsation).
But if the crust crystallizes before {\boldmath $\Omega$}
has time to align with {\boldmath $\mu$}, then $\vv{e}_3$
and {\boldmath $\Omega$} are not necessarily aligned
(depending on the relative size of the crystalline
and pre-accretion magnetic deformation) and
the star does precess.
Moreover, this conclusion does not change when a mountain is
subsequently accreted along {\boldmath $\mu$};
the new $\vv{e}_3$ (nearly, but not exactly, parallel to
{\boldmath $\mu$}) is still misaligned with
{\boldmath $\Omega$} in general.
Gravitational waves are emitted at $f_*$ and $2f_*$.
Of course, internal dissipation after crystallization
(and, indeed, during accretion) may force
{\boldmath $\Omega$} to align with $\vv{e}_3$
(cf. Earth).\footnote{
Accreting millisecond pulsars like
SAX J1808.4$-$3658 do not show evidence of precession
in their pulse shapes, but it is not clear how
stringent the limits are
(Galloway, private communication).
}$^{,}$\footnote{
We do not consider the magnetospheric accretion torque
here \citep{lai99}.
}
If this occurs, the precession stops and
the gravitational wave signal at $f_*$ disappears.
The smaller signal at $2f_*$ persists if the star is triaxial
(almost certainly true for any realistic magnetic mountain,
even though we do not calculate the triaxiality explicitly
in this paper)
but disappears if the star is biaxial (which is unlikely).
To compute the polarization waveforms with precession included,
one may employ the
small-wobble-angle expansion 
for a nearly spherical star derived by \citet{zim80}
and extended to quadratic order by \citet{van05}.
This calculation lies outside the scope of this paper
but constitutes important future work.

Recent coherent, multi-interferometer searches for continuous
gravitational waves from nonaxisymmetric pulsars appear
to have focused on the signal at $2f_*$, to the
exclusion of the signal at $f_*$.
Examples include the S1 science run of the LIGO and GEO 600 detectors,
which was used to place an upper limit $\epsilon\leq 2.9\times 10^{-4}$
on the ellipticity of the radio millisecond pulsar
J1939$+$2134 \citep{lig04a},
and the S2 science run of the three LIGO I detectors
(two 4-km arms and one 2-km arm), which was used to place upper
limits on $\epsilon$ for 28 isolated pulsars with
$f_* > 25$ Hz \citep{lig04b}.
Our results indicate that these (time- and frequency-domain)
search strategies must be revised to include the signal
at $f_*$ (if the mountain is static) and even to collect
signal within a bandwidth $\Delta f$ centered at $f_*$ and $2f_*$
(if the mountain oscillates).
This remains true under several of the evolutionary scenarios
outlined above when precession is included,
depending on the (unknown) competitive balance between driving
and damping.

The analysis in this paper disregards the fact that LIGO II will be
tunable.
It is important to redo the SNR calculations with realistic
tunable noise curves, to investigate whether
the likelihood of detection is maximized by observing near
$f_*$ or $2f_*$.
We also do not consider several physical processes that affect
magnetic burial, such as sinking of accreted material, Ohmic dissipation,
or Hall currents; their importance is estimated roughly by
\citet{mel05}.
Finally,
Doppler shifts due to the Earth's orbit and rotation
\citep[e.g.][]{bon96} are neglected, as are
slow secular drifts in sensitivity during a
coherent integration.


\acknowledgments
{
This research was supported by an
Australian Postgraduate Award.
}
\bibliographystyle{apj}
\bibliography{b}

\begin{thebibliography}{39}
\expandafter\ifx\csname natexlab\endcsname\relax\def\natexlab#1{#1}\fi

\bibitem[{{Abramowitz} \& {Stegun}(1972)}]{abr72}
{Abramowitz}, M. \& {Stegun}, I.~A. 1972, {Handbook of Mathematical Functions}
  (Handbook of Mathematical Functions, New York: Dover, 1972)

\bibitem[{{Alpar} \& {Sauls}(1988)}]{alp88}
{Alpar}, M.~A. \& {Sauls}, J.~A. 1988, \apj, 327, 723

\bibitem[{{Bildsten}(1998)}]{bil98}
{Bildsten}, L. 1998, ApJ, 501, L89

\bibitem[{{Bonazzola} \& {Gourgoulhon}(1996)}]{bon96}
{Bonazzola}, S. \& {Gourgoulhon}, E. 1996, A\&A, 312, 675

\bibitem[{{Brady} {et~al.}(1998){Brady}, {Creighton}, {Cutler}, \&
  {Schutz}}]{bra98}
{Brady}, P.~R., {Creighton}, T., {Cutler}, C., \& {Schutz}, B.~F. 1998, Phys.
  Rev. D, 57, 2101

\bibitem[{{Chakrabarty} {et~al.}(2003){Chakrabarty}, {Morgan}, {Muno},
  {Galloway}, {Wijnands}, {van der Klis}, \& {Markwardt}}]{cha03}
{Chakrabarty}, D., {Morgan}, E.~H., {Muno}, M.~P., {Galloway}, D.~K.,
  {Wijnands}, R., {van der Klis}, M., \& {Markwardt}, C.~B. 2003, Nat, 424, 42

\bibitem[{{Cook} {et~al.}(1994){Cook}, {Shapiro}, \& {Teukolsky}}]{coo94}
{Cook}, G.~B., {Shapiro}, S.~L., \& {Teukolsky}, S.~A. 1994, ApJ, 424, 823

\bibitem[{{Creighton}(2003)}]{cre03}
{Creighton}, T. 2003, Classical and Quantum Gravity, 20, 853

\bibitem[{{Cutler}(2002)}]{cut02}
{Cutler}, C. 2002, Phys. Rev. D, 66, 084025

\bibitem[{{Cutler} \& {Jones}(2001)}]{cut01}
{Cutler}, C. \& {Jones}, D.~I. 2001, Phys. Rev. D, 63, 024002

\bibitem[{{Deshpande} \& {McCulloch}(1996)}]{des96}
{Deshpande}, A.~A. \& {McCulloch}, P.~M. 1996, in ASP Conf. Ser. 105: IAU
  Colloq. 160: Pulsars: Problems and Progress, 101

\bibitem[{{Jaranowski} {et~al.}(1998){Jaranowski}, {Kr{\' o}lak}, \&
  {Schutz}}]{jar98}
{Jaranowski}, P., {Kr{\' o}lak}, A., \& {Schutz}, B.~F. 1998, Phys. Rev. D, 58,
  063001

\bibitem[{{Jones} \& {Andersson}(2002)}]{jon02}
{Jones}, D.~I. \& {Andersson}, N. 2002, MNRAS, 331, 203

\bibitem[{{Katz}(1989)}]{kat89}
{Katz}, J.~I. 1989, MNRAS, 239, 751

\bibitem[{{Lai}(1999)}]{lai99}
{Lai}, D. 1999, ApJ, 524, 1030

\bibitem[{{Link} \& {Epstein}(2001)}]{lin01}
{Link}, B. \& {Epstein}, R.~I. 2001, ApJ, 556, 392

\bibitem[{{Link} {et~al.}(1998){Link}, {Franco}, \& {Epstein}}]{lin98}
{Link}, B., {Franco}, L.~M., \& {Epstein}, R.~I. 1998, ApJ, 508, 838

\bibitem[{{Lyne} {et~al.}(1988){Lyne}, {Pritchard}, \& {Smith}}]{lyn88}
{Lyne}, A.~G., {Pritchard}, R.~S., \& {Smith}, F.~G. 1988, MNRAS, 233, 667

\bibitem[{{Melatos}(2000)}]{mel00a}
{Melatos}, A. 2000, MNRAS, 313, 217

\bibitem[{{Melatos} \& {Payne}(2005)}]{mel05}
{Melatos}, A. \& {Payne}, D. J.~B. 2005, ApJ, 623, 1044

\bibitem[{{Melatos} \& {Phinney}(2001)}]{mel01}
{Melatos}, A. \& {Phinney}, E.~S. 2001, PASA, 18, 421

\bibitem[{{Mouschovias}(1974)}]{mou74}
{Mouschovias}, T. 1974, ApJ, 192, 37

\bibitem[{{Payne} \& {Melatos}(2004)}]{pay04}
{Payne}, D. J.~B. \& {Melatos}, A. 2004, MNRAS, 351, 569

\bibitem[{{Payne} \& {Melatos}(2005)}]{pay05}
---. 2005, MNRAS, submitted

\bibitem[{{Pines} \& {Shaham}(1974)}]{pin74}
{Pines}, D. \& {Shaham}, J. 1974, Comments on Astrophysics and Space Physics,
  6, 37

\bibitem[{{Shabanova} {et~al.}(2001){Shabanova}, {Lyne}, \& {Urama}}]{sha01}
{Shabanova}, T.~V., {Lyne}, A.~G., \& {Urama}, J.~O. 2001, ApJ, 552, 321

\bibitem[{{Shakura} {et~al.}(1998){Shakura}, {Postnov}, \& {Prokhorov}}]{sha98}
{Shakura}, N.~I., {Postnov}, K.~A., \& {Prokhorov}, M.~E. 1998, aa, 331, L37

\bibitem[{{Stairs} {et~al.}(2000){Stairs}, {Lyne}, \& {Shemar}}]{sta00}
{Stairs}, I.~H., {Lyne}, A.~G., \& {Shemar}, S.~L. 2000, Nat, 406, 484

\bibitem[{{Stone} \& {Norman}(1992)}]{sto92}
{Stone}, J.~M. \& {Norman}, M.~L. 1992, ApJS, 80, 753

\bibitem[{{The LIGO Scientific Collaboration: B.~Abbott}
  {et~al.}(2004{\natexlab{a}}){The LIGO Scientific Collaboration: B.~Abbott},
  {Kramer}, \& {Lyne}}]{lig04b}
{The LIGO Scientific Collaboration: B.~Abbott}, {Kramer}, M., \& {Lyne}, A.~G.
  2004{\natexlab{a}}, ArXiv General Relativity and Quantum Cosmology e-prints

\bibitem[{{The LIGO Scientific Collaboration: B.~Abbott}
  {et~al.}(2004{\natexlab{b}}){The LIGO Scientific Collaboration: B.~Abbott},
  {Kramer}, \& {Lyne}}]{lig04a}
---. 2004{\natexlab{b}}, Phys. Rev. D, 69, 082004

\bibitem[{{Thorne}(1987)}]{tho87}
{Thorne}, K.~S. 1987, {Gravitational radiation, in Three Hundred Years of
  Gravitation, edited by W. W. Hawking and W. Israel } (Cambridge, MA:
  Cambridge University Press, 1987), 330--458

\bibitem[{{Ushomirsky} {et~al.}(2000){Ushomirsky}, {Cutler}, \&
  {Bildsten}}]{ush00}
{Ushomirsky}, G., {Cutler}, C., \& {Bildsten}, L. 2000, MNRAS, 319, 902

\bibitem[{{Van Den Broeck}(2005)}]{van05}
{Van Den Broeck}, C. 2005, Classical and Quantum Gravity, 22, 1825

\bibitem[{{Villarreal} \& {Strohmayer}(2004)}]{vil04}
{Villarreal}, A.~R. \& {Strohmayer}, T.~E. 2004, ApJ, 614, L121

\bibitem[{{Wijnands} {et~al.}(2003){Wijnands}, {van der Klis}, {Homan},
  {Chakrabarty}, {Markwardt}, \& {Morgan}}]{wij03b}
{Wijnands}, R., {van der Klis}, M., {Homan}, J., {Chakrabarty}, D.,
  {Markwardt}, C.~B., \& {Morgan}, E.~H. 2003, Nat, 424, 44

\bibitem[{{Wynn} \& {King}(1995)}]{wyn95}
{Wynn}, G.~A. \& {King}, A.~R. 1995, MNRAS, 275, 9

\bibitem[{{Zimmermann}(1980)}]{zim80}
{Zimmermann}, M. 1980, Phys. Rev. D, 21, 891

\bibitem[{{Zimmermann} \& {Szedenits}(1979)}]{zim79}
{Zimmermann}, M. \& {Szedenits}, E. 1979, Phys. Rev. D, 20, 351

\end{thebibliography}



\end{document}